\documentclass[%
 aip,
 jmp,%
 amsmath,amssymb,
reprint,%
]{revtex4-1}

\usepackage{graphicx}
\usepackage{dcolumn}
\usepackage{bm}

\begin{document}


\title{Reduction of critical field for magnetic and orbital-ordering phase transition  
in impurity-substituted Nd$_{0.45}$Sr$_{0.55}$MnO$_3$ crystal}
\author{Y. Izuchi}
\affiliation{Department of Physics, Sophia University, Tokyo 102-8554, Japan}

\author{M. Akaki}
\affiliation{The Institute for Solid State Physics, The University of Tokyo, Kashiwa 277-8581, Japan}

\author{D. Akahoshi}
\affiliation{Department of Physics, Toho University, Funabashi 274-8510, Japan}

\author{H. Kuwahara}
\affiliation{Department of Physics, Sophia University, Tokyo 102-8554, Japan}

\date{\today}

\begin{abstract}
We have investigated the Mn-site substitution effect in Nd$_{0.45}$Sr$_{0.55}$MnO$_3$ single crystal, which has an $A$-type layered antiferromagnetic ($A$-AFM) phase with the 3$d_{x^2-y^2}$-type orbital-order.  
Substitution of Fe or Ga for Mn-site suppresses both the $A$-AFM order and competing ferromagnetic (FM) correlation whereas Cr substitution suppresses only the $A$-AFM order but reactivates the underlying FM correlation via double-exchange mechanism along the AFM coupled $c$-direction.  
In Nd$_{0.45}$Sr$_{0.55}$Mn$_{0.95}$Cr$_{0.05}$O$_3$, the $A$-AFM state with the orbital-order is changed into the orbital-disordered three-dimensional FM metallic state by applying magnetic field of $\mu_0 H = 12$ T, which is much smaller than that of the parent compound Nd$_{0.45}$Sr$_{0.55}$MnO$_3$.
\end{abstract}
%

\maketitle

Perovskite manganites $R_{1-x}A_x$MnO$_3$ ($R$ and $A$ being trivalent rare-earth and divalent alkaline-earth ions, respectively) exhibit rich electronic phases such as charge-ordered (CO) and ferromagnetic (FM) metallic states.\cite{kuwaharaSci,kajimoto,akahoshi}
These electronic phases can be controlled via band-width and/or band-filling, which often leads to drastic phase conversion, i.e., the colossal magnetoresistance (CMR) effect.
Besides band-width and band-filling, quenched disorder arising from substitution of impurity for Mn is another effective method for electronic phase control of $R_{1-x}A_x$MnO$_3$.
The magnetic and electronic properties of $R_{1-x}A_x$MnO$_3$ are quite sensitive to the kind and amount of impurities.
For instance, in Nd$_{0.5}$Ca$_{0.5}$MnO$_3$, which has the CO and $CE$-type antiferromagnetic (AFM) insulating ground state, only 7 \% substitution of Cr for Mn changes the ground state from the CO and $CE$-AFM insulating state to the FM metallic one.\cite{schuddinck,kimura,machida}
Much effort has been devoted to study on the impurity effect on $R_{1-x}A_x$MnO$_3$ with $x \leq 0.5$, but not with $x > 0.5$ because the CMR effect is mainly observed in $R_{1-x}A_x$MnO$_3$ with $x \leq 0.5$.\cite{bernabe,raveau,hebert,NABabushkina,Moritomo,RMahendiran,Mori,TSOrlova}  
Thus, for further understanding of the impurity effect, the impurity effect on the heavily-doped manganites needs to be studied.

In this study, we have systematically investigated the impurity effect on the heavily doped manganite, Nd$_{0.45}$Sr$_{0.55}$Mn$_{0.95}B_{0.05}$O$_3$ ($B$ = Cr, Fe, and Ga) in single crystalline form.
In order to clarify the role of the electronic configurations of dopants, we employed Cr, Fe, and Ga as dopants; the ionic valences of all the dopants are trivalent with the electronic configurations of 3$d^3$ ($t_{2g}^3e_g^0$, $S$ = 3/2), 3$d^5$ ($t_{2g}^3e_g^2$, $S$ = 5/2), and 3$d^{10}$ ($t_{2g}^6e_g^4$, $S$ = 0), respectively.  
Among the perovskite manganite systems, Nd$_{1-x}$Sr$_x$MnO$_3$ (NSMO) with moderate band-width exhibits the rich electronic phase diagram.\cite{kajimoto}
In NSMO with $0.52 \leq x < 0.625$ including the parent compound ($x$=0.55) of this study, the ground state is an $A$-type layered AFM ($A$-AFM) phase with the 3$d_{x^2-y^2}$-type orbital-order, in which each FM metallic layer within the $ab$ plane is antiferromagnetically coupled with its adjacent layer along the $c$ axis.\cite{kuwaharaPRL,kawano}   
Reflecting the magnetic and orbital-ordering transition, the orthorhombic (nearly tetragonal, $Pbnm$) O$^{\ddagger}$ structure with the lattice constants $a\sim b<c/\sqrt{2}$ changes to orthorhombic O$^{\prime}$ structure with $c/\sqrt{2}<b<a$ below $T_{\rm N}$\@. The FM $ab$-plane expands and the AFM coupled $c$-axis shrinks, implying that the 3$d_{x^2-y^2}$-type orbitals lie in the FM $ab$-plane and the transfer interaction along the AFM coupled $c$-direction almost vanishes. Because of its characteristic spin- and orbital-orders, the magneto-transport properties exhibit large anisotropy, and the $A$-AFM phase can be regarded as two-dimensional (2D) ferromagnet.\cite{kuwaharaPRL} 
This $A$-AFM phase competes with the FM metallic phase so that application of magnetic fields induces the transition from the $A$-AFM phase to the FM metallic one (i.e., the 2D FM phase to the 3D FM one).\cite{hayashi}
However, the $A$-AFM order is robust against magnetic fields; in Nd$_{0.45}$Sr$_{0.55}$MnO$_3$, the 2D FM to 3D FM transition occurs at $\mu_0 H = 35$ T\@.  
In this study, we demonstrate that the robustness of the $A$-AFM phase with the orbital-order can be controlled through impurity doping method, leading to a novel gigantic magnetoresistive effect. 

Nd$_{0.45}$Sr$_{0.55}$Mn$_{0.95}B_{0.05}$O$_3$ ($B$ = Cr, Fe, and Ga) single crystals were prepared as follows.  
We first prepared the polycrystalline samples using Nd$_2$O$_3$, SrCO$_3$, Mn$_3$O$_4$, Cr$_2$O$_3$, Fe$_2$O$_3$, and Ga$_2$O$_3$ as starting materials.
The mixed powders with an appropriate molar ratio were calcined at 1273 K for 24 h in air with an intermediate grinding.  
Then, the calcined powders were pressed into a rod shape (5 mm diameter $\times$ 100 mm) and sintered at 1623 K for 12 h in air.  
The single crystalline samples were grown from the sintered rods with a floating zone method in air at a feed rate of $5-10$ mm/h.  
We confirmed that all the grown crystals are of single phase without any traces of impurity phases by temperature-controllable powder X-ray diffraction (XRD) measurements, which evidenced that the substitution of impurity (Cr, Fe, and Ga) for Mn site is completely accomplished without crystallographic defects.  
The single crystalline samples were randomly cut into a rectangular shape with the size larger than a single crystallographic domain size to average the magnetic and transport anisotropies.  
Magnetic properties were measured with the Physical Property Measurement System (PPMS, Quantum Design).  
Magneto-transport properties were measured with a conventional four-terminal method using a temperature-controllable cryostat equipped with a superconducting magnet that can provide a magnetic field up to 12 T\@.

\begin{figure}
\begin{center}
\includegraphics[width=6cm]{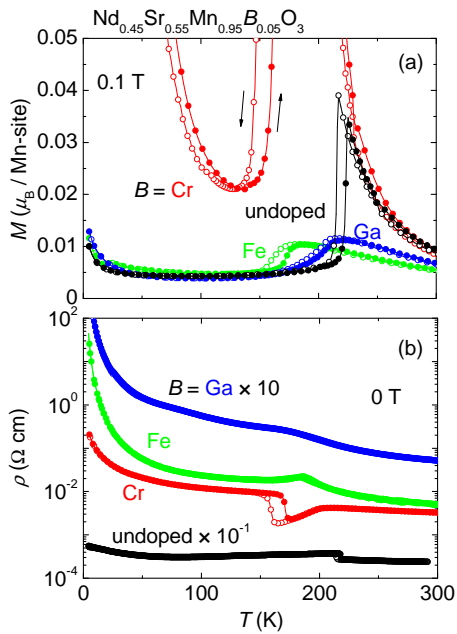}
\end{center}
\vspace*{-5mm}
\caption{Temperature dependence of (a) magnetization in $\mu_0 H = 0.1$ T and (b) resistivity of Nd$_{0.45}$Sr$_{0.55}$Mn$_{0.95}B_{0.05}$O$_3$ ($B$ = Cr, Fe, and Ga) crystals.  
The resistivity of undoped and Ga-doped crystals is shifted vertically for clarity.  
Warming run after zero-field cooling and field cooling run are denoted as closed and open circles, respectively.  
}
\label{fig1r}
\end{figure}

Figure 1 shows the temperature dependence of the magnetization and resistivity of Nd$_{0.45}$Sr$_{0.55}$Mn$_{0.95}B_{0.05}$O$_3$ ($B$ = Cr, Fe, and Ga) single crystals.  
In the pristine compound Nd$_{0.45}$Sr$_{0.55}$MnO$_3$, the magnetization steeply drops at the $A$-AFM transition temperature $T_{\rm N} = 220$ K, accompanying a thermal hysteresis owing to the first-order nature of the transition.
The positive Curie-Weiss temperature $T_{\rm CW}$ (not shown) indicates that the FM and AFM correlations compete with each other above $T_{\rm N}$.\cite{kuwaharaPRL,yoshizawa}
The resistivity shows a small jump at $T_N$, being barely metallic below and above $T_N$.  
These results are consistent with previous reports,\cite{kuwaharaPRL,hayashi} where the anisotropic transport properties and the magnetic-field-induced phenomena of Nd$_{0.45}$Sr$_{0.55}$MnO$_3$ have been already reported in more detail.  
5 \% substitution of Cr for Mn drastically reduces the $T_N$ by $\sim 50$ K; instead the underlying FM state replaces the $A$-AFM state in the temperature range of 170 K - 200 K (see also Fig.\ 2. (a)).  
The positive $T_{\rm CW}$ for Nd$_{0.45}$Sr$_{0.55}$Mn$_{0.95}$Cr$_{0.05}$O$_3$ is almost same as or slightly increased compared with that of the parent compound.  
The FM magnetization at $\mu_0 H = 0.1$ T reaches about 1 $\mu_{\rm B}$/Mn-site immediately above $T_N$.  
The resistivity of Nd$_{0.45}$Sr$_{0.55}$Mn$_{0.95}$Cr$_{0.05}$O$_3$ exhibits an anomaly at the FM transition temperature $T_{\rm C} = 200$ K, and metallic temperature dependence is found in the FM region.
Then the resistivity abruptly jumps at $T_{\rm N} = 170$ K with a thermal hysteresis, below which it shows insulating behavior (see also in Fig.\ 2 (b)).
That is, at this temperature, the 3D FM metallic state with orbital-disorder is quenched, sharply changing into the $A$-AFM state with the 3$d_{x^2-y^2}$-type orbital-order.  
In accordance with these magnetic and electronic transitions, the crystallographic structural change has been confirmed by XRD: The tetragonal-like O$^{\ddagger}$ structure corresponding to the 3D FM metallic phase with orbital-disorder changes to the orthorhombic O$^{\prime}$ with the $A$-AFM non-metallic and 3$d_{x^2-y^2}$-type orbital-ordered phase. In the vicinity of $T_{\rm N}$, we have observed the above two-phase coexisting state, which can be distinguish by Bragg peaks in XRD patterns, which means they possess a crystallographic long range order as well as magnetic one.

The effects of Ga- and Fe-dopings are contrastive to that of Cr-doping.
In Ga-doped Nd$_{0.45}$Sr$_{0.55}$MnO$_3$, the magnetization is considerably suppressed immediately above the $T_{\rm N}$ compared with that of Nd$_{0.45}$Sr$_{0.55}$MnO$_3$, and the $T_{\rm N}$ is slightly lowered to 208 K, as clearly seen in Fig.\ 1. (a).  
The resistivity of Nd$_{0.45}$Sr$_{0.55}$Mn$_{0.95}$Ga$_{0.05}$O$_3$ is much higher than that of the undoped compound, and metallic behavior is not observed in the measured temperature range.  
The curvature of the resistivity slightly changes around $T_{\rm N}$.
Similar to Nd$_{0.45}$Sr$_{0.55}$Mn$_{0.95}$Ga$_{0.05}$O$_3$, in Fe-doped Nd$_{0.45}$Sr$_{0.55}$MnO$_3$, the FM correlation is largely suppressed, and the $T_{\rm N}$ is much reduced down to 171 K, compared with those of the pristine and Ga-doped compounds.
We note that the positive $T_{\rm CW}$ is reduced by 15 and 52 K for the cases of Ga- and Fe-doping, respectively.  
The resistivity of Nd$_{0.45}$Sr$_{0.55}$Mn$_{0.95}$Ga$_{0.05}$O$_3$ is semiconducting in the whole temperature region, exhibiting a broad peak around $T_N$.  
Hereafter, we concentrate on the magnetic-field-induced phenomena observed in Nd$_{0.45}$Sr$_{0.55}$Mn$_{0.95}$Cr$_{0.05}$O$_3$, which is distinct from a simple degradation of the magnetic interactions observed in the Ga- and Fe-doped systems.

\begin{figure}
\begin{center}
\includegraphics[width=6cm]{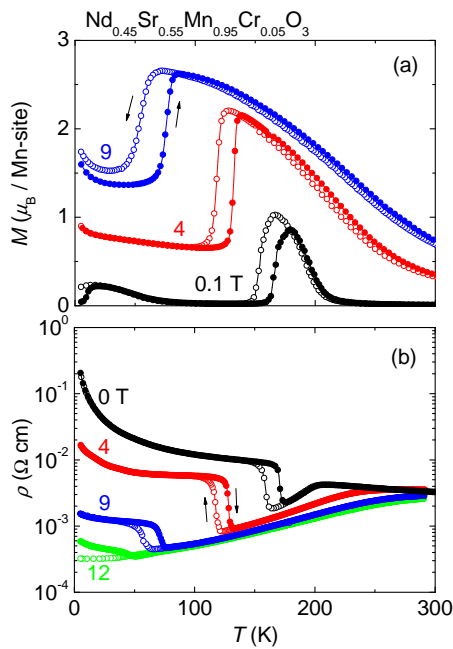}
\end{center}
\vspace*{-5mm}
\caption{Temperature dependence of (a) magnetization and (b) resistivity of Nd$_{0.45}$Sr$_{0.55}$Mn$_{0.95}$Cr$_{0.05}$O$_3$ crystal under several fixed magnetic fields.  
Warming run after zero-field cooling and field cooling run are denoted as closed and open circles, respectively. 
}
\label{fig2r2}
\end{figure}

\begin{figure}
\begin{center}
\includegraphics[width=6cm]{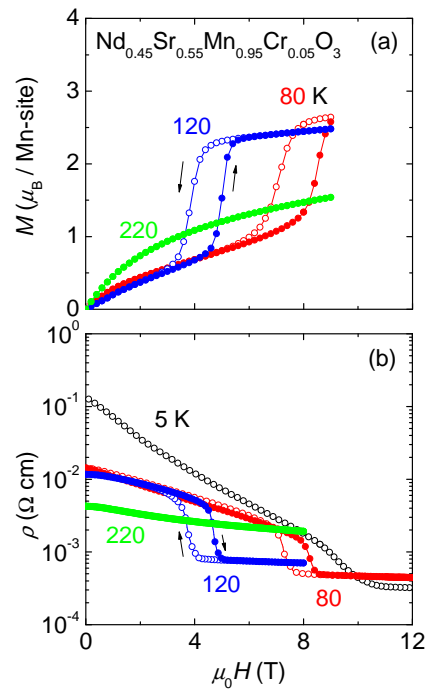}
\end{center}
\vspace*{-5mm}
\caption{Magnetic field dependence of (a) magnetization and (b) resistivity of Nd$_{0.45}$Sr$_{0.55}$Mn$_{0.95}$Cr$_{0.05}$O$_3$ crystal at several fixed temperatures.  
Field increasing and decreasing runs are denoted as closed and open circles, respectively.  
These isothermal $M$-$H$ and $\rho$-$H$ measurements were performed after the sample was cooled down to a target temperature without a magnetic field.  
The only $\rho$-$H$ curve at 5 K was measured in a field decreasing process after the sample was cooled in a field of 12 T\@.}
\label{fig3}
\end{figure}

We show in Fig.\ 2 the temperature dependence of the magnetization and resistivity of Nd$_{0.45}$Sr$_{0.55}$Mn$_{0.95}$Cr$_{0.05}$O$_3$ measured at fixed magnetic fields.  
As already discussed above, in a zero field, the FM state appears below $T_{\rm C} = 200$ K and then turns into the $A$-AFM one at $T_{\rm N} = 170$ K with the concomitant metal to non-metal transition.  
With increasing magnetic fields, the $T_{\rm N}$ linearly shifts to lower temperatures, and in the case of the field cooling process, the $A$-AFM state completely disappears at $\mu_0 H = 12$ T, being replaced by the FM metallic state.
The $M$-$H$ curves and isothermal magnetoresistances of Nd$_{0.45}$Sr$_{0.55}$Mn$_{0.95}$Cr$_{0.05}$O$_3$ are displayed in Figs.\ 3 (a) and 3 (b), respectively.  
The $M$-$H$ curve (at 220 K) near above $T_{\rm C}$ shows a nonlinear increase due to the existence of the FM correlation, and the resistivity exhibits small magnetoresistance.  
At 120 K and 80 K, the sudden increase of the magnetization and the sudden drop of the resistivity, i.e., the large  magnetoresistance are detected with a hysteresis at the respective critical fields.  
The magnetoresistive ratio of $\rho$ (0 T) to $\rho$ (12 T) reaches more than two orders of magnitude at 5 K, as seen from the isothermal scan shown in Fig.\ 3 (b).  
These abrupt changes indicate that application of magnetic fields induces the first-order transition from the $A$-AFM non-metallic state to the FM metallic one below $T_{N} = 170$ K.  
This field-induced phase transition is due to Zeeman energy gain of the FM phase under external magnetic fields, and responsible for the large magnetoresistance.  
Similar magnetic-field-induced phenomena are also observed in the undoped sample, Nd$_{0.45}$Sr$_{0.55}$MnO$_3$, but the critical field of 5 \% Cr-doped Nd$_{0.45}$Sr$_{0.55}$MnO$_3$ is drastically reduced compared with Nd$_{0.45}$Sr$_{0.55}$MnO$_3$.
Small magnetoresistance is also observed in both the paramagnetic and $A$-AFM regions under magnetic fields smaller than the critical fields.  
This small magnetoresistance is explained by reduction of spin scattering of $e_g$-electrons owing to Mn 3$d$ spins forcedly aligned towards the direction of applied magnetic fields, which is commonly observed in perovskite manganites.  


Figure\ 4 shows the phase diagram of Nd$_{0.45}$Sr$_{0.55}$Mn$_{1-y}$Cr$_{y}$O$_3$ ($y = 0$ and 0.05) crystals in the magnetic-field and temperature plane.  
The data of the parent compound Nd$_{0.45}$Sr$_{0.55}$MnO$_3$ ($y = 0$) are also included for comparison.\cite{hayashi}
The phase boundary between the $A$-AFM and FM orders is determined by the isothermal magnetoresistance measurements, which is also consistent with the thermal ones.  
As seen from the phase diagram, the critical field for the AFM to FM phase transition becomes larger with decreasing temperature in both $y = 0$ and 0.05 whereas the critical field of $y = 0.05$ is much smaller than that of $y = 0$.  
At 5 K, the critical magnetic field of $y = 0.05$ is $\mu_0 H = 9.4$ T, which is drastically reduced by about 25 T from that of Nd$_{0.45}$Sr$_{0.55}$MnO$_3$.  


\begin{figure}
\begin{center}
\includegraphics[width=6cm]{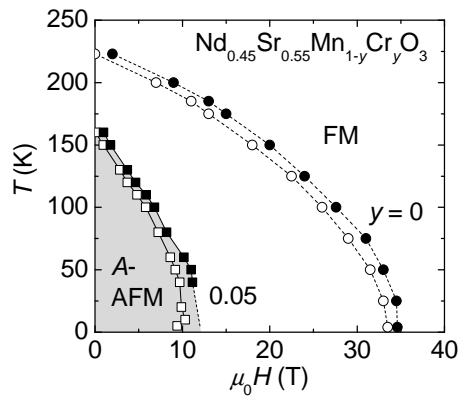}
\end{center}
\vspace*{-5mm}
\caption{The phase diagram in the magnetic-field and temperature plane for Nd$_{0.45}$Sr$_{0.55}$Mn$_{1-y}$Cr$_y$O$_3$ ($y$ = 0, 0.05) crystals.  
The transition from the $A$-type AFM state with orbital-order to FM one with orbital-disorder is denoted as closed symbols while that in the reverse direction as open ones.  
The data for $y$ = 0 sample are taken from Ref. 17.}
\label{fig4}
\end{figure}

As demonstrated above, the impurity effects on the magnetic and transport properties of Nd$_{0.45}$Sr$_{0.55}$MnO$_3$ markedly depend on the dopants, and they can be classified into two groups.  
(1) In the case of Cr-doping, the $A$-AFM interaction is suppressed, while the underlying FM metallic state emerges.  
(2) In the cases of Ga- or Fe-dopings, both the $A$-AFM order and the competing FM correlation are weakened.  
First we discuss the origin of the FM metallic behavior induced by Cr-doping.  
The emergence of the FM correlation is observed only in Cr-doped Nd$_{0.45}$Sr$_{0.55}$MnO$_3$, but not in Ga- and Fe-doped ones.  
Therefore, it is clear that the origin is not attributed to the ionic valence of the dopants, since all the dopants used in this study have the same ionic valence of +3.  
One of the significant differences between these dopants is the electronic configuration: Cr$^{3+}$, Fe$^{3+}$, and Ga$^{3+}$ have the electronic configurations of $t_{2g}^3e_g^0$, $t_{2g}^3e_g^2$, and $t_{2g}^6e_g^4$, respectively.  
Among these dopants, only Cr$^{3+}$ has no $e_g$ electron, suggesting that the $e_g$ orbitals of the dopants play an important role in the emergence of the FM metallic phase observed in Nd$_{0.45}$Sr$_{0.55}$Mn$_{0.95}$Cr$_{0.05}$O$_3$.  
In Cr-doped Nd$_{0.45}$Sr$_{0.55}$MnO$_3$, since the $e_g$ orbitals of Cr$^{3+}$ is empty, the $e_g$ electrons of Mn$^{3+}$ can travel through Cr$^{3+}$ sites, which have the electronic configuration equivalent to that of Mn$^{4+}$, mediating the FM double-exchange interaction along the $c$ direction. It is consistent with the results of resistivity and XRD measurements.  
In contrast to the maintaining of the FM correlation, the $A$-AFM order is suppressed by the quenched disorder arising from impurity substitution as evidenced by decreasing $T_N$.  
Therefore, the decrease of AFM interaction would reactivate the FM correlation hidden in the AFM state.
According to the recent papers (Refs. 21-24), the microscopic origin of the FM interaction induced by substitution of Cr for Mn site is now still open question under debate: Both of superexchange and double-exchange interactions are proposed to explain the FM interaction between Cr$^{3+}$ and Mn$^{3+}$. As for our experimental data shown in the figures 1 and 2, metallic conduction appears concomitantly with the FM state. This result strongly supports that FM interaction is due to the double-exchange interaction which is mediated by mobile $e_g$ electrons. The mobile carriers are confined within the FM $ab$-plane and cannot move along the AFM coupled $c$-direction in the parent Cr-undoped manganite. By contrast, the carriers in the Cr-doped one can move along the $c$-direction through around the Cr sites where the three-dimensional orbital mixing due to local lattice distortion is occurred and the hopping along the $c$-direction is revived.
As already mentioned, we have confirmed the tetragonal-like O$^{\ddagger}$ structure in the FM metallic region by XRD. The 3D FM orbital-disordered metallic clusters are attributed to the above-mentioned revival of the double-exchange interaction along the AFM coupled $c$-direction.

In the case of Ga$^{3+}$, the $e_g$ orbitals are fully occupied so that the $e_g$ electrons of Mn$^{3+}$ cannot move through Ga$^{3+}$ sites, that is, the FM double-exchange interaction does not work.  
On the other hand, the ionic state of Fe ions is rather complicated, since Fe ions in perovskite oxides can take the ionic valences of both +3 and +4.
Arima $et$ $al$. reported that the charge-transfer (CT) gaps of LaFeO$_3$ and LaMnO$_3$ are 2 eV and 1 eV, respectively.\cite{arima}
This result indicates that Fe$^{3+}$ is more stable than Mn$^{3+}$ in the perovskite structure and that the $e_g$ electrons of Fe$^{3+}$ tend to be localized.  
Thus, since there is no vacancy on the up-spin band of the $e_g$ orbital of Fe$^{3+}$, the $e_g$ electrons of Mn$^{3+}$ cannot move through Fe$^{3+}$ sites.  
Consequently, FM double-exchange interaction is not effective like the case of Cr-doping.
For this reason, substitution of Ga$^{3+}$ and Fe$^{3+}$ suppresses the FM correlation while that of Cr$^{3+}$ does not.  

We have investigated the impurity effect of Nd$_{0.45}$Sr$_{0.55}$Mn$_{0.95}B_{0.05}$O$_3$ ($B$ = Cr, Fe, and Ga) single crystals and have found that the impurity effect strongly depends on the electronic configuration of the dopants.  
In the cases of Fe- and Ga-dopings, both the $A$-AFM state and the competing FM correlation are suppressed due to the quenched disorder inherent in impurity substitution.
On the other hand, substitution of Cr$^{3+}$ would reactivate the FM correlation due to the double-exchange interaction hidden in the $A$-AFM order which is substantially suppressed by the quenched disorder.
Consequently, Cr-doping drastically reduces the critical magnetic field from the $A$-AFM phase with the 3$d_{x^2-y^2}$-type orbital-order to the FM metallic one with orbital-disorder, i.e., the 2D FM state to the 3D FM one. 
Fine modification of crystal structure other than band-width and band-filling, i.e. $B$-site-substitution demonstrated here and/or oxygen isotope methods,\cite{NABabushkina} would open up novel paths to synthesize functional oxide materials.

\bigskip

We would like to thank T. Goto and Y. Shindo for their help in measuring magnetoresistance and S. Mashimo for growing a single crystal.  
This work was supported by JSPS KAKENHI Grant Numbers 25420174, 24540383, and partly supported by Grant-in-Aid for JSPS Fellows from Japan Society for Promotion of Science.

\end{document}